\documentclass[showpacs,aps]{revtex4}
\usepackage{graphicx,psfrag}
\usepackage{color}
\def\bc{\begin{center}}
\def\ec{\end{center}}

\def\beq{\begin{equation}}
\def\eeq{\end{equation}}

\def\beq{\begin{equation}}
\def\eeq{\end{equation}}

\def\br{{\bf r}}
\def\bk{{\bf k}}
\def\bq{{\bf q}}
\def\bp{{\bf p}}

\begin{document}
\title{Sensitive linear response of an electron-hole superfluid in a periodic potential}

\author{Oleg L. Berman$^{1,2}$, Roman Ya. Kezerashvili$^{1,2}$,  Yurii E.
Lozovik$^{3}$, and Klaus Ziegler$^{1,4}$}
\affiliation{\mbox{$^{1}$Physics Department, New York City College
of Technology, The City University of New York,} \\
Brooklyn, NY 11201, USA \\
\mbox{$^{2}$The Graduate School and University Center, The
City University of New York,} \\
New York, NY 10016, USA \\
\mbox{$^{3}$Institute of Spectroscopy, Russian Academy of
Sciences,} \\
142190 Troitsk, Moscow, Russia \\
\mbox{$^{4}$   Institut f\"ur Physik, Universit\"at Augsburg\\
D-86135 Augsburg, Germany }}
\date{\today}

\begin{abstract}
We consider excitons in a two-dimensional periodic potential and
study the linear response of the excitonic superfluid to an
electromagnetic wave at low and high densities. It turns out that
the static structure factor for small wavevectors is very sensitive
to a change of density and temperature. It is a
consequence of the fact that thermal fluctuations play a crucial
role at small wavevectors, since exchanging the order of the two
limits, zero temperature and vanishing wavevector, leads to different results for
the structure factor. This effect could be used for high accuracy measurements in the
superfluid exciton phase, which might be realized by a gated
electron-hole gas. The transition of the exciton system
from the superfluid state to a non-superfluid state and its manifestation
by light scattering are discussed.
\end{abstract}


\maketitle

\section{Introduction}
\label{intro}

A gas of excitons is a canonical example for an interacting
many-body system \cite{Snoke_review}. Excitons are bound states of
an electron and a hole, which can be considered as a system with no
effective electric charge. This justifies the assumption of a
short-range interaction. The latter is caused by the Pauli principle
because of the fermionic constituents. At low temperatures the
excitons can form a superfluid as a collective many-body state.
Moreover, in the presence of a periodic potential a Mott-insulating
(MI) state is possible where the excitons form a state which is
commensurate to the periodic structure of the potential minima. Both
states, the superfluid and the MI phase, and the phase transition
between them have been observed in an ultracold gase of bosonic
rubidium atoms \cite{greiner02}. Although it is more difficult to
control the parameters of an exciton gas with present technologi
   es, the similarity of the exciton gas and a bosonic gas of real
atoms indicates that the formation of these two states and the phase
transition between them should be accessible under proper
experimental conditions.

Superfluidity in a two-dimensional (2D) system of electron-hole
pairs was predicted on the basis of Cooper pairing within a BCS
mean-field approach~\cite{Schrieffer} in
Ref.~\cite{Lozovik}. The BCS phase of electron-hole Cooper pairs in
a dense electron-hole system and a dilute gas of indirect excitons,
formed as bound states of electron-hole pairs, were also analyzed in
coupled quantum wells (CQW)~\cite{LY,BLSC}. This was followed by a
number of detailed
theoretical~\cite{Shevchenko,Lerner,Dzjubenko,Kallin,Knox,Yoshioka,Littlewood,Vignale,Ulloa}
as well as experimental
studies~\cite{Snoke_paper,Snoke_paper_Sc,Chemla,Krivolapchuk,Timofeev,Zrenner,Sivan,Snoke,EM}.
Besides the superfluid phase, a Wigner supersolid state  caused by
dipolar repulsion in electron-hole bilayers, was
described~\cite{JBD}. The recent theoretical and experimental
achievements in the studies of the superfluid dipolar exciton phases
in CQWs  were reviewed in Ref.~\cite{Snoke_review},
and various experimental studies of excitonic phases in CQWs were described in Ref.~\cite{Butov_JPCM}.

The formation of a superfluid with indirect excitons was recently proposed in Ref. \cite{fogler14}
using MoS$_2$ layers separated by a hexagonal boron nitride insulating barrier and surrounded by
hexagonal boron nitride cladding layers and in Ref.~\cite{berman16} for dipolar excitions
in two parallel transition metal dichalcogenides layers. A perpendicular electric field
modifies the band structure in a way that it becomes advantageous for optically excited
electrons and holes to reside in the opposite MoS$_2$ monolayers to form indirect excitons.
As an extension of these proposals, we consider excitons in a somewhat modified design, using a
structured gate to create an additional periodic potential, to study the linear response of the
excitonic superfluid to an electromagnetic wave at low and high concentrations.
This provide a new controllable scale through the length of the periodicity.


Besides the question about how to create collective states in an
exciton gas it is also important to characterize and analyze these
states experimentally. Although the system is assumed to be
translational invariant, there are characteristic spatial
correlations of quantum fluctuations. Here we suggest to measure the
linear response to an external electromagnetic field. This response
depends on the wavevector $\bq$ of the external field, which
provides information related to the spatial properties of the state.
Using the static structure factor (SSF), the response is related to
the spatial correlation function of the exciton density
\cite{pitaevskii}. In this paper we will study the  SSF inside the
superfluid phase and how it changes when we get close to the phase
boundaries in the dilute regime and in the high density regime
(i.e., near the transition to an MI phase).

The paper is organized in the following way. In Sec. \ref{sect:model}, starting from an
interacting gas of electrons and holes,
we derive an effective model for excitons formed by bound electron-hole states.
Assuming Coulomb interaction, we neglect the possibility for a dissociation of electrons and holes.
This model is treated in mean-field approximation and the Green's function of quasiparticles are
calculated in Subsect. \ref{sect:mfa_quasi}. In Sec. \ref{sect:thermal} we study the corresponding
SSF at non-zero temperature and compare it with the result of zero temperature.
The results are discussed for different regimes in Sec. \ref{sect:discussion}. Conclusions follow
in Sec. V.

\section{Model of a electron-hole gas in an external periodic potential}
\label{sect:model}

The Hamiltonian of a the system of electrons and holes, either in single layer or spatially separated
in a double layer, can be written in momentum representation as
 \beq
H=\sum_{\bp}\sum_{\sigma=e,h}(\epsilon_{\bp,\sigma}-\mu_\sigma)c^\dagger_{\bp\sigma}c_{\bp\sigma}
+\sum_{\bp,\bp_{1},\bp_{2}} U_\bp
c^\dagger_{\bp-\bp_{1},h}c_{\bp-\bp_{2},h}c^\dagger_{\bp_{1},e}c_{\bp_{2},e}
\ , \label{hamilton} \eeq where $c^\dagger_{p,e}$ ($c_{p,e}$) is the
creation (annihilation)  operator for electrons, and
$c^\dagger_{p,h}$ ($c_{p,h}$) are the corresponding operators for
holes. While $\mu_1$ is the chemical potential of elctrons, $\mu_2$ is the chemical potential
of holes, assuming the concentrations of
electrons and holes are equal in order to have a neutral electron-hole plasma. This is justified since the
electrons and holes are created always pairwise by an external laser
source.  The electron and hole single-particle energy spectra $\epsilon_{\bp,\sigma}$ are defined
in a tight-binding approximation, reflecting an external periodic field which is applied to the system.
The electron-hole attraction due to Coulomb interaction is given in momentum space as $U_\bp$,
which is finite for $q\sim 0$ as a result of screening. Its specific form is different whether we
have direct or indirect excitions. In Eq.~(\ref{hamilton})
the spins of electrons and holes are neglected, because we are not interested in magnetization effects.

The fermionic Hamiltonian could be treated within the BCS mean-field approximation to obtain
the order parameter for Cooper pairs \cite{Schrieffer}
For many physical quantities,
such as the superfluid concentration, this approximation is sufficient. Correlations of excitons, though, would
also require to include quantum fluctuations generated by quasiparticles. There are several options
for including quantum excitonic effects. A very direct way is to consider an effective Hamiltonian $H_b$
which describes the dynamics of excitons on the lattice, based on tunneling of excitons
between neighboring lattice sites.
When we assume that the effect of the Coulomb interaction is strong, for
instance, in case of a small distance between electrons in one and holes in another layer,
the excitons will not dissociate into electrons and holes. Then we can consider the excitons as
stable quantum particles, where the
shortest distance of the model is the lattice spacing, because the excitons tunnel between the
neighboring minima of the periodic potential. The fact that excitons are formed by fermions leads to
a local repulsive interaction. Similar to the famous Hubbard model, the competition of nearest
neighbor tunneling on the lattice and the on-site repulsion can lead to the rich physics of
``strong correlations'' with different quantum states, including superfluidity, Mott and topological states.
This requires the tuning of the tunneling rate and the concentration of particles. Both parameters
are tunable in our experimental proposal through the variation of an external electric field.

\subsection{Effective exciton model}
\label{sect:mfa_quasi}

First, the idea is to derive from the fermionic Hamiltonian (\ref{hamilton}) an effective
Hamiltonian for bound electron-hole pairs (excitons) whose creation operators in configuration
space can be written as
\beq
a_{\br}^\dagger=\int_{\br',\br''}K_{\br-\br',\br-\br''}c_{\br',h}^\dagger c_{\br'',e}^\dagger
d^2r' d^2r''
\ ,
\label{exc_op1}
\eeq
where the kernel $K_{\br-\br',\br-\br''}$ decays exponentially on a characteristic length $\xi$
away from $\br$. This length represents the effective size of the exciton.
Now we assume that the lattice spacing is much larger than $\xi$, which implies that the
exciton creation and annihilation operators are in a good approximation local:
\beq
a_\br^\dagger=c_{\br,h}^\dagger c_{\br,e}^\dagger \ , \ \ \ a_\br=c_{\br,e}c_{\br,h}
\ .
\label{exc_op2}
\eeq
These operators resemble hard-core bosons, since they satisfy the bosonic commutation relations with the
additional condition ${a_\br^\dagger}^2=0$ due to the Pauli principle of the fermionic constituents.

Second, we assume that the excitons cannot dissociate into fermions.
Therefore, the effective Hamiltonian is expressed by the operators $a_\br$ and $a_\br^\dagger$ only.
As a first order approximation we can write
\beq
H_{eff}=\sum_{\br,\br'}(J_{\br,\br'}+\mu\delta_{\br,\br'})a_\br^\dagger a_{\br'}
\ ,
\label{eff_ham00}
\eeq
where the summation is with respect to the minima of the periodic potential,
using a tight-binding approximation. In Eq. (\ref{eff_ham00}) the hopping rate is
\beq
J_{\br,\br'}=\cases{
J, &  $\br,\br'$ nearest neighbors \cr
0, & otherwise \cr
}
\eeq
describes the tunneling of excitons between neighboring potential minima,
while $\mu$ is the chemical potential of excitons that controls their concentration. This Hamiltonian
describes essentially the physics of the exciton gas in a periodic potential with on-site interaction.
Its extension to other types of interaction, including long-range dipole-dipole
interaction, is obvious but not important for the subsequent discussion.

Hamiltonian (\ref{eff_ham00}) can have a normal and a superfluid state for the exciton gas.
In particular, the MI state with concentration $n_{\rm tot}=1$ is an eigenstate of $H_{eff}$.
Considering that the system is at temperatures  well below the
Kosterlitz-Thouless transition temperature, one can use a mean-field
approximation and follow the idea of the Bogoliubov approach. Then
the superfluid phase is distinguished from the normal phase by a
spatially uniform order parameter $\phi=|\phi|e^{i\alpha}$, which
vanishes outside the superfluid phase. Then the exciton operators in
the superfluid phase read \beq a_\br=\phi+\varphi_\br \ ,\ \ \
a_\br^\dagger=\phi^*+\varphi_\br^\dagger \ , \label{operators} \eeq
where $\varphi_\br^\dagger$ is the creation operator for bosonic
quasiparticles. If $N_0$ is the number of bosons in the superfluid
phase and $N$ is the number of lattice sites, $n_0=N_0/N$ is the
concentration of superfluid excitons per unit cell of the lattice.
$N_0$ is obtained from the order
parameter as $N_0=|\phi|^2$. A mean-field approximation for the
Hamiltonian (\ref{eff_ham00}) gives at zero temperature
\cite{moseley07,moseley08}
\begin{equation}
 n_0 = \left\{ \begin{array}{l@{\quad}l}
 \left(1-\mu^2/J^2\right)/4, & \mbox{if } -1<\mu/J<1 \\
 0, & \mbox{otherwise} \end{array} \right.  .
 \label{n0}
\end{equation}
Moreover, the total excitonic concentration $n_{\rm tot}=N_{\rm
tot}/N$ per unit cell of the lattice is obtained from the free
energy per site \beq F=-\frac{1}{\beta N}\log\left({\rm Tr}
e^{-\beta H_b}\right) \eeq as
\begin{equation}
 n_{\rm tot} =\frac{\partial F}{\partial\mu} \sim\left\{ \begin{array}{l@{\quad\mbox{if }}l}
 0, & \mu/J\le-1 \\
 \left(1+\mu/ J\right)/2, & -1< \mu/J< 1 \\
 1, & \mu/J\ge1 \end{array} \right.
 \ .
 \label{ntot}
\end{equation}
Therefore, concentrations are measured in units of the lattice constant and
appear in our calculation as dimensionless quantities.

\begin{figure*}[t]
\includegraphics[width=9cm]{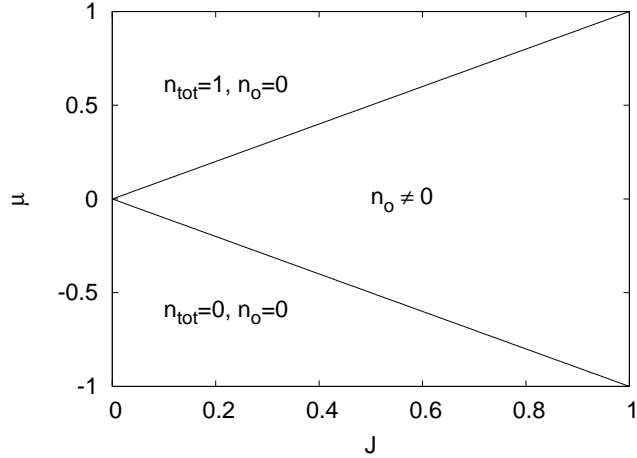}
\caption{
Excitonic phase diagram at $T=0$ with three different phases, separated by two second-order transition lines
(from Eq. (\ref{n0}), in arbitrary units). The lower phase
is a MI phase with $n_{\rm tot}=0$, $n_0=0$, the upper phase is a MI phase with $n_{\rm tot}=1$, $n_0=0$
and the intermediate phase is a superfluid phase with $0<n_{\rm tot}<1$, $n_0>0$. The concentrations along a
vertical cut through the intermediate phase at a fixed hopping rate $J$ are plotted in Fig. \ref{fig:densities}.
}
\label{fig:phased}
\end{figure*}

\begin{figure*}[t]
\includegraphics[width=9cm]{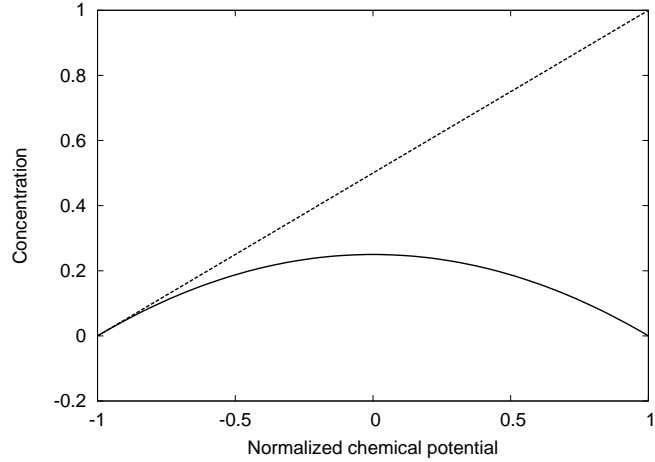}
\caption{
The total excitonic concentration $n_{\rm tot}$ (dashed curve) and the condensate concentration $n_0$ (solid curve)
in mean-field approximation (cf. Eqs. (\ref{n0}) and (\ref{ntot})) for $-1\le\mu/J\le 1$, where
the normalized chemical potential is $\mu/J$.
Lattice effects prevent that all excitons participate in the superfluid as the concentration is increasing,
indicating a crossover to a MI state.
}
\label{fig:densities}
\end{figure*}

In Fig. \ref{fig:densities} the total and the superfluid
concentration are plotted in the superfluid phase, shown in Fig.
\ref{fig:phased}, as a function of the chemical potential. In the
dilute regime, that corresponds to  low
total concentration, almost all excitons are in the superfluid
phase, whereas a reduction of the superfluid
concentration with increasing total concentration is caused by the
formation of a kind of supersolid due to the interplay of the
interaction and the underlying lattice structure.

The rather simple results in Eqs. (\ref{n0}) and (\ref{ntot}) can be used to calculate other physical
quantities of the excitonic gas. For instance, the effect of quantum fluctuations in momentum representation
is described by the
quasiparticle Green's function of $\delta\phi_\bq={\tilde\varphi}_\bq+{\tilde\varphi}_\bq^\dagger$
\cite{ziegler94,moseley07,moseley08}:
\beq
\langle\delta{\phi}_{\bq,\omega}\delta{\phi}_{-\bq,\omega} \rangle
=\frac{(1-\kappa_\bq^2)(1+1/\kappa_\bq)}{(1-\kappa_\bq^2)[2n_0-1/2+1/(1+\kappa_\bq)] -(\hbar\omega/J)^2}
\ ,
\label{GF1}
\eeq
where $\kappa_\bq$ is the dispersion in the periodic potential, normalized with $J$, and $\omega$ is the frequency
of the external electromagnetic field.
This result is similar to the two-particle Green's function obtained for the exciton gas in Ref. \cite{keldysh68}.

\section{Static structure factor}
\label{structure}

The response of a many-body quantum system to a weak external electromagnetic field of frequency $\omega$
and wavevector $\bq$ is given by Fermi's Golden Rule, where the latter is characterized by the
structure factor $S(\bq,\omega)$ \cite{caupin08,pitaevskii}. This quantity is defined as the
Fourier transform of the truncated density-density correlation function:
\beq
S(\bq,\omega)=\frac{1}{2\pi n_{\rm tot}}\int_{\bk,\bp}\int\left[
\langle {\tilde a}_\bk(t) {\tilde a}^\dagger_{-\bk-\bq}(t){\tilde a}_\bp(0) {\tilde a}^\dagger_{-\bp-\bq}(0)\rangle
-\langle {\tilde a}_\bk(t) {\tilde a}^\dagger_{-\bk-\bq}(t)\rangle
\langle{\tilde a}_\bp(0) {\tilde a}^\dagger_{-\bp-\bq}(0)\rangle
\right] e^{i\omega t}dt d^2p d^2k
\ ,
\eeq
where the wavevector $\bq$ is measured in units of the inverse lattice constant $1/a$.
Using the truncated correlation function (i.e., substracting the product of concentrations by the second
term) eliminates a Dirac delta function at $\bq=0$.

With the operators in Eq. (\ref{operators}), $\phi=\sqrt{N_0}$ and the quasiparticle Green's function
(\ref{GF1}) we obtain 
\beq
S(\bq,\omega)=\frac{n_0}{n_{\rm tot}}\langle\delta{\phi}_{\bq,\omega}\delta{\phi}_{-\bq,\omega} \rangle
\ .
\eeq
Integration over all frequencies $\omega$ gives us the SSF as
\beq
S(\bq)\equiv\frac{1}{2\pi}\int S(\bq,\omega)d\omega
=\frac{n_0}{4n_{\rm tot}}\frac{(1-\kappa_\bq^2)(1+1/\kappa_\bq)}{\sqrt{(1-\kappa_\bq^2)[2n_0-1/2+1/(1+\kappa_\bq)]}}
\label{static_sf2}
\eeq
with the lattice tight-binding dispersion $\kappa_\bq$. In the case of
a square lattice with lattice constant $a$ we get $\kappa_\bq=(\cos a q_1+\cos a q_2)/2$.
Replacing $J(1-\kappa_\bq)$ by the dispersion of a free Bose gas $\hbar^2 q^2/2m$ and assuming
that $\hbar^2 q^2/2m\ll J$, we obtain from Eq. (\ref{static_sf2})
the well-known result of Feynman for the low concentration regime $n_0/n_{\rm tot}\sim 1$ \cite{feynman54}:
\beq
S(\bq)\sim \frac{\hbar^2q^2/2m}{\sqrt{(\hbar^2 q^2/2m)[4Jn_0+\hbar^2q^2/2m]}}
\ .
\eeq

\subsection{Thermal fluctuations}
\label{sect:thermal}

The SSF at zero temperature is characterized by a
vanishing behavior for a vanishing wavevector $\bq$.
The question is now whether or not this behavior is also visible in
a realistic experiment where we have thermal fluctuations. We will
demonstrate subsequently that this is not the case, since even very
small thermal fluctuations alter the behavior of the SSF
substantially at small wavevectors.

\begin{figure*}[t]
\includegraphics[width=9cm]{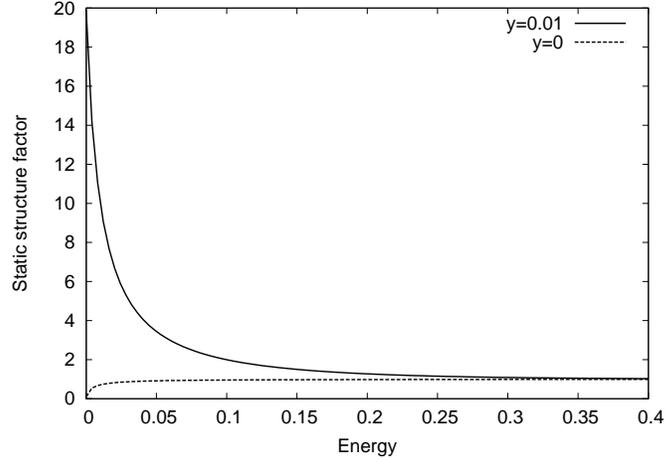}
\caption{
Dimensionless SSF as a function of the dispersion energy $Jg_\bq$ for $n_0=0.0025$ in units
of inverse unit cell area and different temperatures:
$\beta J=0$ (dashed curve) and $\beta J=0.01$ (solid curve). At zero temperature the SSF
vanishes for $\bq\to0$.
}
\label{fig:struct}
\end{figure*}

The calculation for including thermal fluctuations requires a summation with respect to Matsubara
frequencies $\omega_l=2\pi l/\beta\hbar$, where $l=0,1,2,...$ \cite{abrikosov}, rather than a frequency integration.
This summation can be expressed again by an integral in the complex plane \cite{negele},
which gives us for the superfluid phase \cite{pitaevskii,moseley07,moseley08}
\begin{equation}
S(\bq)\sim
\frac{n_0}{n_{\rm tot}}\frac{g_{\bq}}{\epsilon_{\bq}}\coth\frac{\beta J\epsilon_{\bq}}{2}
\ \ {\rm with}\ \
\epsilon_\bq=\sqrt{4n_0g_\bq+(1-4n_0)g_\bq^2}
\label{structurf2}
\end{equation}
and with the bare dispersion $g_\bq=1-(\cos a q_1+\cos a q_2)/2$ for the square lattice.
The SSF is very sensitive to thermal fluctuation at small $\bq$,
as illustrated in Fig. \ref{fig:struct} for $n_0=0.0025$: It vanishes
for $T=0$ and $\bq=0$, whereas it increases strongly for $\beta J=0.01$ if $\bq\sim 0$.
The reason for this behavior is the $\coth x$ term, which
diverges for $x\sim0$. This factor is 1 for strictly zero temperature but diverges for any
finite $\beta$ when $\bq\sim 0$.

\begin{figure*}[t]
\includegraphics[width=9cm]{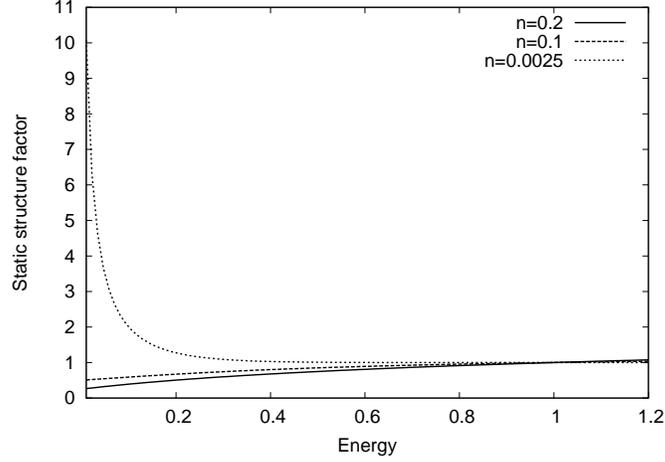}
\caption{
Dimensionless SSF as a function of the dispersion energy $g_\bq$ in units of $J$ for $\beta J=0.01$ and
different superfluid concentrations: $n_0=0.2$ (solid curve), $n_0=0.1$ (dashed curve), $n_0=0.0025$ (dotted curve).
}
\label{fig:struct2}
\end{figure*}

\section{Discussion}
\label{sect:discussion}


Let us distinguish three different regimes for the SSF: $\mu\sim -J$
(low concentration regime), $\mu=0$ (maximum superfluid concentration) and
$\mu\sim J$ (vanishing superfluid at high concentration). For the behavior
of the SSF it is crucial that the limit $T\to0$ followed by the limit
$\bq\to0$ leads to a different result for the structure factor than
the limit $\bq\to0$ followed by the limit $T\to0$.  Therefore,
we have to be careful with the regime $\bq\sim0$ at low temperature,
because the SSF is very sensitive to small changes of parameters.
This can be seen in Figs. \ref{fig:struct}, \ref{fig:struct2}, where
either a small change in temperature or a small change in the superfluid
concentration has a strong effect on the SSF.


{\it i. Low concentration regime}
In the dilute regime with $\mu\sim -J$ we have $n_0/n_{\rm tot}=1$, such that
we obtain from Eq. (\ref{structurf2})
\beq
S(\bq)\sim \frac{g_{\bq}}{\epsilon_{\bq}}\coth\frac{\beta J\epsilon_{\bq}}{2}
\sim \coth\frac{\beta J\epsilon_{\bq}}{2}
\ ,
\label{sf_dilute1}
\eeq
which is in agreement with the well-known result for the weakly interacting Bose gas
and with the SSF of the ideal (noninteracting) Bose gas  \cite{pitaevskii}.
For low concentration, when $\beta J\epsilon_{\bq}\ll 1$ one can obtain from Eq. (\ref{sf_dilute1})
\beq
S(\bq)\sim\frac{2}{\beta J g_\bq}
\ .
\label{sf_dilute}
\eeq
Interaction between the excitons can have two effects: it can create a superfluid with concentration $n_0>0$ and,
together with the periodic potential, it can destroy the superfluid and create a MI state. Thus,
besides the weakly interacting regime with $n_{\rm tot}\sim n_0\sim0$, there is an intermediate regime with
$n_{\rm tot}\sim 0.5$ and maximum superfluid concentration $n_0\sim 0.25$, and a strongly interacting regime of
a dense exciton gas with $n_{\rm tot}\sim 1$ and $n_0\sim0$. Since $n_0$ is small at low and at high concentration
(cf. Fig. \ref{fig:densities}), the SSF differs only by the prefactor $n_0/n_{\rm tot}$ in
these two regimes, which is 1 in the dilute regime and very small in the dense regime.

{\it ii. Maximum superfluid concentration}
In this case we have $\mu=0$, which implies according to Eqs. (\ref{n0}) and (\ref{ntot})
$n_0=1/4$ and $n_{\rm tot}=1/2$ (half-filled lattice); i.e., $n_0/n_{\rm tot}=1/2$.
Moreover, we get for the quasiparticle dispersion $\epsilon_\bq=\sqrt{g_\bq}$ and, therefore,
\beq
S(\bq)\sim \frac{1}{2}\frac{g_{\bq}}{\epsilon_{\bq}}\coth\frac{\beta J\epsilon_{\bq}}{2}
\sim \frac{1}{2}\sqrt{g_\bq}\coth\frac{\beta J\sqrt{g_\bq}}{2}
\sim\frac{1}{\beta J}\ \ {\rm for}\ \beta J\sqrt{g_\bq}\ll 1
\ .
\label{asympt2}
\eeq

{\it iii. High concentration regime}
$\mu\sim J$ implies for the concentrations $n_0\sim 0$ and $n_{\rm tot}\sim 1$; i.e., $n_0/n_{\rm tot}\sim 0$.
Thus, in the dense regime, i.e. close to the MI phase when $n_{\rm tot}\approx 1$,
the SSF vanishes with $n_0$ as
\beq
S(\bq)\sim n_0\frac{g_{\bq}}{\epsilon_{\bq}}\coth\frac{\beta J\epsilon_{\bq}}{2}
\sim\frac{1}{2}\frac{1}{1+g_\bq/4n_0}\frac{1}{\beta J}\ \ {\rm for}\ \beta J\epsilon_{\bq}\ll 1
\ ,
\label{asympt3}
\eeq
in contrast to Eq. (\ref{sf_dilute}) for the dilute regime.

\begin{figure*}[t]
\includegraphics[width=9cm]{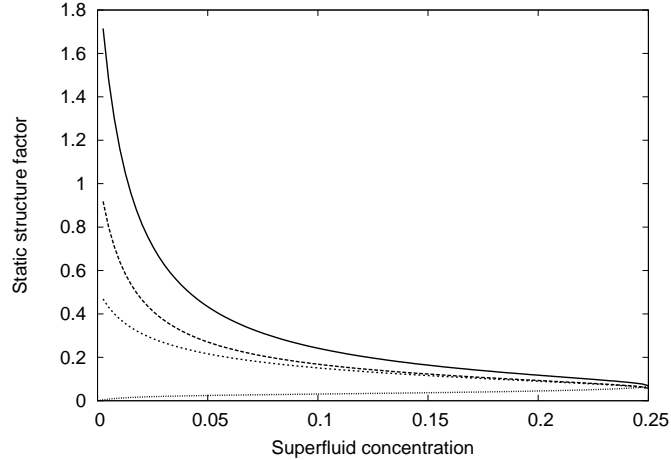}
\caption{
Dimensionless SSF as a function of the superfluid concentration $n_0$ for $g_\bq=0.05$ and different temperatures:
$\beta J=10$, $\beta J=20$, $\beta J=80$ in the low concentration regime $0\le n_{\rm tot}\le 0.5$ and
for $\beta J=10$ in the high concentration regime $0.5\le n_{\rm tot}\le 1$ (from top to bottom).
}
\label{fig:struct3}
\end{figure*}

The dependence of the SSF on $g_\bq$ in the dilute regime is depicted in Fig. \ref{fig:struct}.
At fixed temperature the SSF is plotted in Fig. \ref{fig:struct2} for different values of $n_0$.
Both figures reflect the asymptotic behavior for small $\bq$ obtained in Eqs. (\ref{sf_dilute}),
(\ref{asympt2}) and (\ref{asympt3}).
In the experiment with a dilute exciton gas we should always see a broad maximum at small $\bq$ which decays
according Eq. (\ref{sf_dilute}) like $q^{-2}$. This maximum disappears at higher exciton concentrations,
and the SSF is relatively flat with a very weak increase, as illustrated in
Fig. \ref{fig:struct2}.

We have not calculated the SSF in the MI phase because the latter is not accessible
with the present experimental techniques. Such a calculation would require a different
approach to obtain the corresponding quasiparticle Green's function. But this is a straightforward task
when we use the concept developed, for instance, in Ref. \cite{moseley07}.

\section{Conclusions}

We predict a method to control the state of an exciton system by
introducing a
spatially periodic potential through a profiled external gate or by a
periodic superlattice structure. The control can be performed also
by change of the chemical potential of the system, driven by laser
pumping.  The transition of superfluid (BEC) state to localized
state takes place. This phase transition is controlled by the
parameters of the external potential. The transition can be revealed
by study of exciton flow induced by a gradient of the exciton
concentration, originated from a photon pumping spot. Besides, the
transition can be observed by elastic light scattering. The elastic
light scattering cross section is proportional to  the static
structure factor. We demonstrated that the static structure factor
drastically changes at the superfluid to localized states
transition.

The SSF, which can be measured by non-resonant scattering of electromagnetic
waves, is a useful quantity to characterize a superfluid state in an excitonic system. Since the latter
was proposed some time ago \cite{EM} and observed in a recent experiment \cite{alloing14},
a detailed analysis, for instance, based on X-ray Raman scattering (XRS) \cite{feng07}
could reveal more details of the nature of this state.
In particular, there is a pronounced maximum of the SSF at small wavevectors in the case of a dilute
exciton gas and an almost flat SSF at higher exciton concentrations. This high sensitivity can be employed
for accurate measurements of the superfluid properties at small wavevector $\bq$. In particular,
it would be possible to observe the reduction of the superfluid concentration and the crossover to
the MI phase under an increasing total exciton concentration. In a proper experimental
set-up the latter could be controlled by an external gate.


\end{document}